\documentclass[twocolumn,superscriptaddress,amsmath,aps,prb]{revtex4-1}
\usepackage[utf8]{inputenc}
\usepackage[final]{graphicx}
\usepackage{bm}
\usepackage[normalem]{ulem}
\usepackage[usenames]{color}
\usepackage{times}
\usepackage{verbatim}
\usepackage{xcolor}
\usepackage{hyperref}
\usepackage{wrapfig}
\usepackage{subfigure}
\DeclareGraphicsExtensions{.eps}

\begin{document}
\title{Exceptional points in Fermi liquids with quadrupolar interactions}
\author{Rui Aquino}
\affiliation{Departamento de F{\'\i}sica Te\'orica,
Universidade do Estado do Rio de Janeiro, Rua S\~ao Francisco Xavier 524, 20550-013  
Rio de Janeiro, RJ,  Brazil}
\author{Daniel G. Barci}
\affiliation{Departamento de F{\'\i}sica Te\'orica,
Universidade do Estado do Rio de Janeiro, Rua S\~ao Francisco Xavier 524, 20550-013  
Rio de Janeiro, RJ, Brazil}
\date{\today}

\begin{abstract}
We show the existence of non-Hermitian degeneracies, known as  {\em exceptional points},  in the collective mode spectrum of Fermi liquids with quadrupolar interactions. Through a careful analysis of the analytic properties of the  dynamic quadrupolar susceptibility, we show  that, in the weak attractive region, two stable collective modes coalesce to an exceptional point. We completely characterize this singularity, explicitly showing its topological properties.  Experimental signatures are also discussed. 
\end{abstract}

\maketitle

{\em Introduction--}
Open quantum systems play a central role in most applications of quantum mechanics~\cite{Heinz2002}.  
An important theoretical tool to describe dissipative quantum systems is the modelling of locally non-conservative systems by effective non-Hermitian Hamiltonians~\cite{Rotter2015,Yoshida2018,Michishita2020}. These types of  Hamiltonians has several counterintuitive  properties. Perhaps, one of the most streaking ones is the appearance of  non-Hermitian degeneracies~\cite{Berry1998} known as  {\em exceptional points} (EP)~\cite{Heiss1990,Kato1995}.

When a non-Hermitian Hamiltonian continuously depends on external parameters, it could  happen that, for certain values of the parameters, two or more eigenvalues  coalesce to an EP.  However, this is not a usual degeneracy, as  observed in Hermitian systems. In an EP, not only the eigenvalues coincide but also the eigenvectors become linearly dependent~\cite{Rotter2007}, reducing in this way the dimension of the subspace associated to the degenerated eigenvalue. This singularity of the Hilbert space has remarkable topological consequences~\cite{Yoshida2019, Kawabata2019,Kawabata2019-2,Wojcik2020,kozii2017}. The relation between EPs and dynamical phase transitions was early recognized in theoretical as well as experimental works~\cite{Heiss1998,Rotter1999,Pastawski2006,Eleuch2016}.

In recent years, exciting  findings of EPs  are shown up in very different contexts, strengthening the broad interest of this subject; 
from nuclear~\cite{Kaiser2002} and  atomic physics~\cite{Cartarius2007,Moiseyev2016,Pan2019} to Bose-Einstein condensates \cite{Aharonyan2019} and strongly correlated fermion systems\cite{Yoshida2020}, passing through microwave cavities~\cite{Heiss2001-2}  and SWAP gates in spin systems~\cite{Pastawski2006,Pastawski2008}.
Moreover, topological properties of EPs were experimentally studied in metamaterials set ups~\cite{Kang2017, Park2020}.

In this letter, we report the existence of exceptional points in the  spectrum of collective excitations of Fermi liquids~\cite{nozieres-1999} with higher order Landau parameter interactions.  Fermi liquids with quadrupolar interactions began to call the attention of the condensed matter community because it is the simplest model supporting an isotropic-nematic transition~\cite{OgKiFr2001}. Nematic fluctuations play a crucial role in several strongly correlated systems, such as cuprates  and Fe-based superconductors and a variety of Quantum Hall Effects~\cite{FradKiv2010}.

Collective excitations of Fermi liquids with quadrupolar interactions have been  studied  in different regimes~\cite{OgKiFr2001,Lawler2006,BaRe2013,Rui2019,Chubukov2019,Gochan2020}. 
Here, we explicitly show the appearance of a non-Hermitian singularity for weak quadrupolar attraction. We completely characterize this exceptional point, by analyzing the Hilbert space structure and its topological properties. Finally, we discuss some possible experimental set ups.    
  
{\em Model--}  We consider the simplest  model of  bi-dimensional spinless Fermions with local quadrupolar interactions.  The Hamiltonian is
\begin{align}
H&=\int d^2r\left\{  \psi^\dagger({\bf r}) \epsilon({\bf \nabla}) \psi({\bf r})+\frac{F_2}{4} {\rm Tr}\left[ Q^2({\bf r})\right]\right\}
\label{eq:H}
\end{align}
where $\psi(r)$ is a spinless Fermionic field operator. The bare dispersion relation is given by $\epsilon({\bf \nabla})$, where $\nabla$ is the two-dimensional gradient operator. $F_2$ is the quadrupolar coupling constant. The quadrupolar Fermionic density $Q_{ij}=\psi^\dagger({\bf r})\left[\nabla_i\nabla_j- (\delta_{ij}/2)  \nabla^2\right]\psi({\bf r})$, with $i=1,2$, is a symmetric traceless tensor of rank 2,  invariant under $\pi$ rotations.

Collective modes are encoded in the  dynamic quadrupolar susceptibility (DQS) $\chi_{ijlm}(\omega,{\bf q})=\langle Q_{ij}(-\omega,-{\bf q})Q_{lm}(\omega,{\bf q})\rangle$.  DQS have been intensively studied~\cite{OgKiFr2001,Lawler2006,Zyuzin2018,Chubukov2018-1} in the vicinity of a quantum critical point, where non-Fermi liquid behavior is espected. Conversely, in this letter we study the dynamic response in the Fermi liquid regime. Since the quadrupolar moment has  two  degrees of freedom, the susceptibility has essentially two independent polarizations, the longitudinal  $\chi_2^+(\omega,{\bf q})$ and the transversal polarization $\chi_2^-(\omega,{\bf q})$. These quantities have been computed using different approximation approaches~\cite{OgKiFr2001,Lawler2006,CastroNeto2005,BaRe2013,Rui2019}.
In the limit of small momentum $q<<k_F$, where $k_F$ is the Fermi momentum, the result is~\cite{Lawler2006} (please, see supplemental material for a detailed description of the calculation)
\begin{equation}
  \chi_2^{\pm}(\omega,{\bf q})=\frac{\chi^0_0(s) \pm \chi^0_{4}(s)}{
    1-F_{2}\big(\chi^0_0(s) \pm \chi^0_{4}(s)\big)}
\label{eq:chi2}
\end{equation}
where
\begin{equation}
  \chi^0_{2\ell} =\bigg[-\delta_{\ell,0} +
    K_0(s)\bigg(\frac{1-K_0(s)}{1+K_0(s)}\bigg)^\ell\bigg]
    \label{eq:chi0}
\end{equation}
with  $K_0(s)=s/\sqrt{s^2-1}$. Eq.~(\ref{eq:chi0}) with $\ell=0,2$ are the bare density and quadrupolar susceptibilities, respectively.  Eq~(\ref{eq:chi2}) has the usual structure of an effective interaction in the traditional RPA approximation. Due to the locality of the quadrupolar interaction ({\em i.\ e.\ }, $F_2$ do not depend on ${\bf q}$), the DQS is not a function of $\omega$ and ${\bf q}$ independently. Instead, it depends on the  dimensionless variable $s=\omega/q v_F$, where $\omega$ is the frequency and $q v_F$ is the maximum energy of a particle-hole excitation with momentum $q=|{\bf q}|$ and Fermi velocity $v_F=|{\bf v_F}|$. It is worth mentioning that in the computation of Eq.~(\ref{eq:chi2}), rotational invariance and particle-hole symmetry were imposed.

\begin{figure}[hbt]
\begin{center}
\includegraphics[width=0.38\textwidth]{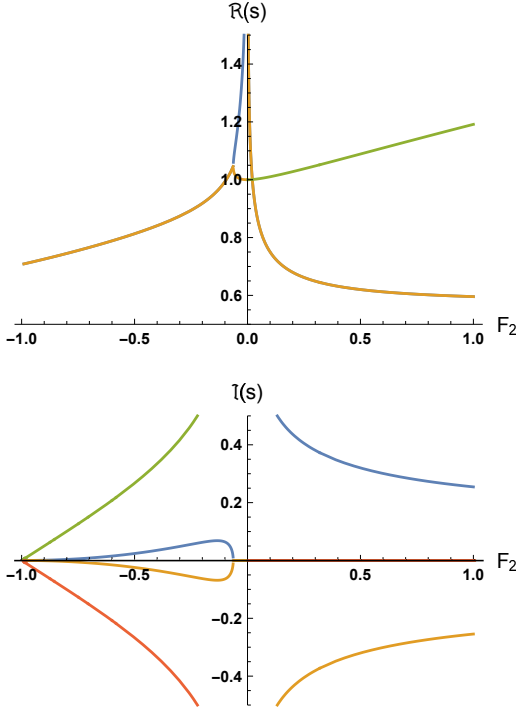}
\end{center}
\caption{Collective modes from the longitudinal polarized component of the DQS $\chi_2^+(s)$. 
In the upper panel we plot $\Re[s(F_2)]$ while in the lower panel we depict $\Im[s(F_2)]$.}
\label{fig:cm}
\end{figure}

{\em Collective modes--}  
The DQS is an analytic function of $s$, having poles and cuts. It has branch points at $s=\pm 1$; the threshold of Landau damping $\omega=\pm v_F q$. We will focus on the longitudinal polarization $\chi_2^+(s)$ since, as we will show,  this component displays an EP. Collective modes are computed by solving the algebraic equation $F_2\left(\chi^0_0(s) +\chi^0_{4}(s)\right)=1$. We have numerically solved it for $F_2$ running  from the strong attractive ($F_2=-1$) to the strong repulsive regime ($F_2>1$).  We display the result in Fig.~(\ref{fig:cm}).
In the upper panel, we show the real part of the collective modes as function of $F_2$, while in lower panel, we show the imaginary part. In the repulsive region ($F_2>0$), we observe a stable (real) mode that tends to $s=1$ when $F_2\to 0$. This is the quadrupolar equivalent of the Landau zero sound. In addition, a damped mode also appears in the same region. The stable mode is continuously extended to the weak attractive region $F_2\lesssim 0$. However, in this regime, there is another stable mode with a divergent behavior,  $s\to +\infty$ when $F_2\to 0^-$. The existence of such a mode was reported in Ref. \onlinecite{Rui2019}. Interestingly, there is a special  point, $F_2^c$, where both stable modes meet together. For $F_2<F_2^c$, these modes become damped as can be clearly seen in the lower panel of Fig. \ref{fig:cm}.  
We can also observe an overdamped mode (purely imaginary) in all the attractive region. This mode is the precursor of the isotropic-nematic phase transition that occurs at $F_2=-1$ and has already been extensively studied~\cite{OgKiFr2001,Lawler2006}.

\begin{figure}[hbt]
\begin{center}
\includegraphics[width=0.38\textwidth]{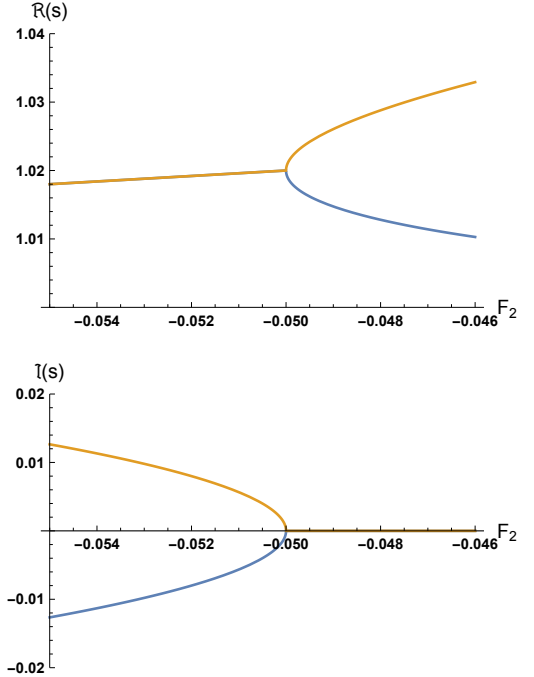}
\end{center}
\caption{Solutions of ${\cal L}^+(s_\pm)=0$, given by Eq.  (\ref{eq:spm}) as a function of the parameter $F_2$.  The upper  panel shows the real part of $s_{\pm}$, while lower one depicts the imaginary part. The point $F_2=-1/20$, where both eigenvalues coalesce and the imaginary part emerges, is the exceptional point.}
\label{fig:EP}
\end{figure}

{\em Exceptional point--} In order to analytically characterize the singularity  at $F_2=F_2^c$, we first observe that $s(F_2^c)\gtrsim 1$, being  well separated from the cut $s^2<1$. On the other hand, the singularity is sufficiently close to $s=1$, allowing us to try a series expansion of $\chi_2^+(s)$ in the neighborhood of $s=1$.  For simplicity, let us work  with the inverse of the DQS, ${\cal L}^+(s)=(\chi_2^+(s))^{-1}$. Expanding this quantity in terms of the variable $\sqrt{(s-1)/2}$, we  find the following expansion (please, see supplemental material for details of the calculation), 
\begin{equation}
{\cal L}^+(s)=-F_2+\sqrt{\frac{s-1}{2}}+5\left(\frac{ s-1}{2}\right)+O\left((s-1)^{3/2}\right).
\label{eq:L1}
\end{equation}
Longitudinal quadrupolar fluctuations $\delta Q^+(s,{\bf q})$ are governed by the effective action
\begin{equation}
S_{\rm eff}=\int \frac{d\omega d^2q}{(2\pi)^3}   {\cal L}^+(s) |\delta Q^+(s,{\bf q})|^2 .
\label{eq:S}
\end{equation}
The collective modes are given by  the roots of ${\cal L}^+(s)=0$. Using Eq. (\ref{eq:L1}), we  obtain
\begin{equation}
s_\pm=\frac{1}{25}\left\{\left(26+10 F_2\right)\pm \sqrt{20 F_2+1}\right\}\; .
\label{eq:spm}
\end{equation} 
$s_{\pm}(F_2)$ have a square root singularity (branch point) at $F_2^c=-1/20$. At this point, both zeros are degenerated, $s_\pm(F_2^c)=51/50$.
We depict the real and imaginary part of $s_\pm(F_2)$ in Fig.~(\ref{fig:EP}).
Thus, the approximation made in Eq.~(\ref{eq:L1}) for $|s-1|<< 1$,  correctly captures the  presence  of the degeneracy point observed in the numerical computation of Fig.~(\ref{fig:cm}). The square-root singularity  is a typical signature of an exceptional point~\cite{Heiss1991}. 

The dynamics described by Eq.~(\ref{eq:L1}) is non-local in time.  
However, since the degeneracy is separated from the cut, we can further expand ${\cal L}^+(s)$ in the neighborhood of $s=s_\pm$.  
In addition, we observe that the local character of the interaction imposes that ${\cal L}^+(s)$ only depends on the dimensionless variable $s$. Thus, we can consider  quadrupolar fluctuations, $\delta Q^+(s)$, ignoring any momentum dependence not scaling with $s$. The consequence is that all collective modes in this approximation have linear dispersion relation $\omega\sim v_F q$. This is a good approximation for weak interactions. However, it breaks down in the  strongly attractive regime ($F_2\sim -1$),  where non-local interactions $F_2(q)$ are essential~\cite{Lawler2006}.  
With these considerations, we arrive to the effective action
\begin{equation}
S_{\rm eff}=\int ds\;  \left\{ \left(s-\epsilon_1\right)\left(s-\epsilon_2\right)+w^2\right\}|\delta Q^+(s)|^2
\label{eq:S1}
\end{equation}
where $\epsilon_1=(1/25)(27+10F_2)$, $\epsilon_2=(1/25)(25+10F_2)$ and $w=(1/25)\sqrt{20 |F_2|}$ are real positive numbers in the vicinity of the EP.  The zeros of the Lagrangian are given of course by  Eq.~(\ref{eq:spm}).

In order to rewrite the effective action in the Hamiltonian formalism  (first order in time), we introduce a two-component vector field $\delta Q^+=(\delta Q_1, \delta Q _2)$.  In terms of this field, the effective action reads  (please, see supplemental material for details)
\begin{equation}
S_{\rm eff}=\int ds\;  (\delta Q^+)^\dagger\left( s I- H_{\rm eff}\right )\delta Q^+   
\label{eq:Seff}
\end{equation}
where  $I$ is the $2\times 2$ identity matrix and the effective Hamiltonian is 
\begin{equation}
H_{\rm eff}=\left(
\begin{array}{cc}
\epsilon_1 & iw  \\
iw & \epsilon_2
\end{array}
\right)
\end{equation} 
It is straightforward to verify that, integrating out the vector component $\delta Q_2$, we obtain the effective action of Eq.~(\ref{eq:S1}) for the field $\delta Q_1$. Therefore, the dynamics near the singularity is driven by a $2\times 2$ symmetric  effective Hamiltonian (non-Hermitian), which determines the properties of the EP~\cite{Eleuch2015-1}.
 
{\em Hilbert space and topology--} The Hilbert space spanned by the basis $\psi_{\pm}$ and its dual, spanned by $\phi_\pm$, are in general different in non-Hermitian Hamiltonian systems. They are defined by 
\begin{align}
H_{\rm eff} \psi_{\pm}&=s_{\pm} \psi_{\pm} 
\label{eq:Hpsi} \\
H^{\dagger}_{\rm eff} \phi_{\pm}&=s^*_{\pm} \phi_{\pm} \; .
\label{eq:Hphi}
\end{align} 
 Bi-orthogonality requires $\langle\phi_i|\psi_j\rangle=\delta_{ij}$ with $i,j=\pm$. 
 Since the effective Hamiltonian is symmetric, the dual space is spanned by  $\phi_\pm=\psi_\pm^*$.
Solving Eq.~(\ref{eq:Hpsi}), we find
\begin{equation}
\psi_\pm= c^\pm \left(
\begin{array}{c}
1  \\
\frac{-i}{\sqrt{1-z}}\left[1\mp z^{1/2}\right]
\end{array}
\label{eq:psi}
\right)\; ,
\end{equation}
where $c^\pm$ are complex normalization constants. We have introduced the variable  $z=1+20 F_2$, in order to have the EP at $z=0$.  As anticipated, not only $s_+=s_-$ at the EP, but the  eigenvectors collapse to $\psi^{\rm EP}_\pm=c^\pm\left(1,-i\right)$. This fact produces that $\langle\phi^{\rm EP}|\psi^{\rm EP}\rangle=0$, which is evidently in conflict with bi-orthogonality. In this way, the EP is a singularity in the structure of the Hilbert space~\cite{Rotter2007}. 
This singularity induces remarkable topological properties. To show this, let us compute the geometric phase that the wave function picks up when the EP is winded in parameter space. For this, we analytically continue $z$ to the complex plane and define 
the Berry phase as $\gamma=i \oint_C d{\bf\ell}\cdot {\bf A}$, where the one-form
${\bf A}= \langle \phi_+|{\bf \nabla}\psi_+\rangle/\langle \phi_+|\psi_+\rangle$
~\cite{Garrison1988},  $C$ is a closed path  and $\nabla$ is the gradient in parameter space $z$. The equivalent definition with $\phi_-$ and $\psi_-$ eigenvectors provides the same result.
 Notice that ${\bf A}$ is ill-defined at the EP since, at this point, the denominator is zero. 
The particular structure of the Hilbert space and its dual allows us to rewrite the vector form as a total derivative (locally a pure gauge), $ {\bf A}= (1/2)\nabla\ln\langle \phi_+|\psi_+\rangle$. Thus, the EP   is a branch point of the logarithm. Each time 
the phase of $\langle \phi_+|\psi_+\rangle$ winds the branch point, the logarithm picks up  a $2\pi i$ term. This property does not depend on the specific path, provided the path encircles the EP. Thus, we can compute $\gamma$ considering a  very small circumference around the EP.  Using Eq.~(\ref{eq:psi}), we find for $|z|<<1$, $\langle \phi_+|\psi_+\rangle\sim z^{1/2}$. Due to the square-root singularity, the phase of   $\langle \phi_+|\psi_+\rangle$ is half the phase of $z$. Therefore, taken the path $C$ winding  {\em two times} the EP, the Berry phase $\gamma=\pi$, in agreement with results obtained for general symmetric non-Hermitian Hamiltonians~\cite{Mailybaev2005}. In this way, in encircling the EP, it is necessary to wind four times the singularity to return to the original state~\cite{Heiss1999}. Recently, this unique  topology of EPs was experimentally confirmed  in metamaterials set ups~\cite{Kang2017, Park2020}. 

\begin{figure}[hbt]
\begin{center}
\includegraphics[width=0.38\textwidth]{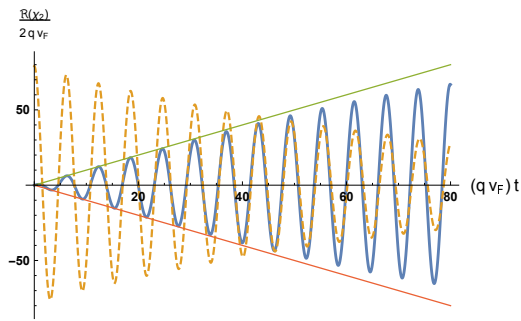}
\end{center}
\caption{ $\Re[\chi_2({\bf q}, t)]/2 v_F q$ as a function of $(v_F q)  t$. The continuous line is plotted with Eq. (\ref{eq:chi-t}) by  fixing  $F_2=F_2^c+0.005$. The dashed line is the damped mode for $F_2=F_2^c-0.005$. The linear functions are the exact modulating  function at the EP, $F_2=F_2^c=-1/20$.}
\label{fig:chi-time}
\end{figure}

{\em Experimental signatures--}  
Information about collective excitations of strongly correlated systems can be obtained by measuring 
momentum-resolved dynamic susceptibility in the meV scale~\cite{Abbamonte2017}. The detection of a stable mode near the usual zero sound could be an indication of the presence of an EP. 
Moreover, pump-probe spectroscopy~\cite{Giannetti2016,Misha2009,Mitrano2019} yield important information on the dynamic response in the time domain. An experimental signature can be obtained from $\chi^+_2({\bf q},t)$, by  Fourier transforming the DQS in the neighborhood of the EP. For $F_2> F_2^c$, the retarded susceptibility is 
\begin{align}
\Re\left[\chi^+_2({\bf q},t)\right]=2v_F q \left[\frac{\sin\left(\omega_- t\right)  }{\omega_-}\right]
\cos\left(\omega_+ t\right) \Theta(t)
\label{eq:chi-t}
\end{align}
where $\omega_\pm=\left(s_+\pm s_-\right)v_F q/2$ and  $\Theta(t)$ is the Heaviside distribution.  We clearly observe two well separated time scales since $\omega_+ / \omega_->>1$.  At the EP, $\omega_-=0$ and $\sin(\omega_- t)/\omega_-\to t$. Thus, the signature of the EP is a growing linear modulating function of time, $\chi^+_2({\bf q},t)\sim t \cos(\omega_+ t)$. An approximate linear modulation can be observed on a huge range of intermediate times, even when the coupling is not fine tuned at  $F_2=F_2^c$. On the other hand, for $F_2<F_2^c$, the dynamic response dramatically changes since the modulation is exponentially damped
$\chi^+_2({\bf q},t)\sim \exp\left\{-|\omega_-| t \right\}\cos(\omega_+ t)$.
We depict these different regimes in Fig. \ref{fig:chi-time}.   
The abrupt change in the  dynamical response at the EP should also be captured in quantum quench set ups~\cite{Iucci2014}. Another interesting possibility is to look for signatures on the AC electrical conductivity\cite{Sodemann2020}.  

{\em Summary and discussion--} We have shown the existence of an EP in the collective mode spectrum of a Fermi liquid with weak attractive quadrupolar interactions. We completely characterize this singularity in terms of the Hilbert space structure as well as through its topological properties. We have also provided experimental signatures in the dynamical response. 
More complex models of Fermi liquids could lead to higher dimensional singularities, such as exceptional lines or surfaces~\cite{Zhen2015,Zhang2019}. For instance, if we consider isotropic density interactions ($F_0$) in addition to the quadrupolar ones\cite{Rui2019}, we still find square-root sigularities which, in the limit of small $F_0$, take the form $s_+-s_-=\sqrt{1+20F_2+4F_0}$. In this way, the spectrum has an {\em exceptional line} parametrized by $F_2+F_0/5=-1/20$.  

Concluding, non-Hermitian singularities appear in the spectrum of collective modes of Fermi liquids with higher angular momentum attractive interactions. Specific properties, such us the singularity location and dimensionality,  are model dependent. However, its existence,  its topological properties and experimental signatures are robust results. It could be important to investigate the influence of these singularities in the single quasi-particle spectrum and its effect on charge transport and other out-of-equilibrium properties.

\acknowledgments
We would like to acknowledge Zochil Gonz\'alez Arenas for  useful comments.
The Brazilian agencies, {\em Funda\c c\~ao de Amparo \`a Pesquisa do Rio
de Janeiro} (FAPERJ), {\em Conselho Nacional de Desenvolvimento Cient\'\i
fico e Tecnol\'ogico} (CNPq) and {\em Coordena\c c\~ao  de Aperfei\c coamento de Pessoal de N\'\i vel Superior}  (CAPES) - Finance Code 001,  are acknowledged  for partial financial support.
RA was partially supported by a  PhD Fellowship from FAPERJ. 


%

\onecolumngrid

\pagebreak

\setcounter{equation}{0}
\setcounter{figure}{0}
\setcounter{table}{0}
\setcounter{page}{1}
\makeatletter
\renewcommand{\theequation}{S\arabic{equation}}
\renewcommand{\thefigure}{S\arabic{figure}}
\renewcommand{\bibnumfmt}[1]{[S#1]}
\renewcommand{\citenumfont}[1]{S#1}

\begin{center}
\textbf{\large Supplemental Material to   ``Exceptional points in Fermi liquids with quadrupolar interactions''}
\end{center}

\section{Dynamic Quadrupolar susceptibility}
In this section we give details of the calculation of the dynamic quadrupolar susceptibility (DQS) in a Fermionic model with local quadrupolar interactions. We begin by considering  the simplest  model of  bi-dimensional spinless Fermions.   The Hamiltonian is
\begin{align}
H&=\int d^2r\left\{  \psi^\dagger({\bf r}) \epsilon({\bf \nabla}) \psi({\bf r})+\frac{F_2}{4} {\rm Tr}\left[ Q^2({\bf r})\right]\right\}
\label{eq:H}
\end{align}
where $\psi(r)$ is a spinless Fermionic field operator. The bare dispersion relation is given by $\epsilon({\bf \nabla})$, where $\nabla$ is the two-dimensional gradient operator. $F_2$ is the quadrupolar coupling constant. The quadrupolar Fermionic density 
\begin{equation}
Q_{ij}({\bf r})=\psi^\dagger({\bf r})\left[\nabla_i\nabla_j- \frac{\delta_{ij}}{2}  \nabla^2\right]\psi({\bf r})
\end{equation}
is a symmetric traceless tensor of rank 2;  $i,j=1,2$.

The dynamic quadrupolar susceptibility (DQS) is given by 
\begin{equation}
\chi_{ijlm}(\omega,{\bf q})=\langle Q_{ij}(-\omega,-{\bf q})Q_{lm}(\omega,{\bf q})\rangle\; .
\label{eq:chi}
\end{equation}
However, the quadrupolar tensor has only two independent degrees of freedom, that can be parametrized by $m_2^+$ and 
$m_2^-$ in the following way
\begin{equation}
Q= \left(
\begin{array}{cc}
m_2^+ & m_2^- \\
m_2^- & -m_2^+ 
\end{array}
\right)
\label{eq:Q}
\end{equation}
Thus, among the sixteen components of Eq. (\ref{eq:chi}), only three are independent, say
\begin{align}
\chi_2^+(\omega,{\bf q})&= \langle m_2^+(-\omega,-{\bf q})m_2^+(\omega,{\bf q})\rangle  \\ 
\chi_2^-(\omega,{\bf q})&=\langle m_2^-(-\omega,-{\bf q})m_2^-(\omega,{\bf q})\rangle  \\
\chi_2^{+-}(\omega,{\bf q})&= \langle m_2^+(-\omega,-{\bf q})m_2^-(\omega,{\bf q})\rangle 
\end{align}

Since the interaction is quartic in the Fermionic fields, the expectations values cannot be computed exactly. 
A computation using the traditional RPA approximation can be found in Ref. [\onlinecite{OgKiFr2001}]. In this paper, we describe the calculation using multidimensional bosonization, originally presented in Ref. [\onlinecite{Lawler2006}]. 
The latter approximation is accurate in a regime of small momentum $|{\bf q}|<<k_F$,  where $k_F$ is the Fermi momentum.  
The result coincides with the former RPA approximation.

The essence of the bosonization approximation is to  describe the effective dynamic  of the system in a regime of energy and momentum very near the Fermi surface. In this sense,  it is a low temperature ($k_B T<< \epsilon_F$, where $\epsilon_F$ is the Fermi energy) and 
long distance  ($q << k_F$) approximation. For weak interactions, this is the same regime of  the Landau theory of Fermi liquids, then, it should reproduce the same results. The advantage of bosonization is that it is not limited to weak interactions. 

It is by now well established that the dynamics induced the the Hamilotian of Eq. (\ref{eq:H})  very near the  Fermi surface can be entirely written in terms of a set of bosonic operators 
\begin{equation}
\delta n_S({\bf r})=:\psi_S^\dagger({\bf r}) \psi_S({\bf r}):
\label{eq:ns}
\end{equation}  
with $S=1,\ldots,N$.  In this equation, $S$ labels the patch in which the Fermi surface was divided, and $\psi_S$ is the Fermionic operator with support on each small patch in momentum space; {\em i.\  e.\  }, $\psi_S(q)$ with $q <\lambda<\Lambda<<k_S$. 
The patch $S$ is a small rectangle centered at ${\bf k}_F={\bf k}_S$ of width $\Lambda$ and height $\lambda$. The colons in Eq. (\ref{eq:ns}) means normal order with respect to a reference state with all the states inside the Fermi surface filled. 
At the end of the calculations the Fermi surface is reconstructed by taking the limit $\Lambda\to 0$, $N\to \infty$, with $N\Lambda=2\pi k_F$.   Within each patch, the Fermionic dispersion relation is linearized, $\epsilon({\bf k})=\epsilon({\bf k}_S)+{\bf v}_S\cdot {\bf q}$, where ${\bf v_S}$ is the Fermi velocity in each patch.  This approximation is a good one in the Fermi liquid phase. However, near the Pomeranchuk regime, it should be corrected by curvature terms. 

It is well known\cite{Lawler2006} that, the effective action of the system can be written in terms of the  set of  density fluctuations $\delta n_S(\omega, {\bf q})$. It can be split  into a free and an interaction part,  $S=S_0+S_I$.
The free term is given by 
\begin{equation}
S_0=\int \frac{d^2 qd\omega}{(2\pi)^3} \sum_S\;\chi^{-1}_S(\omega,{\bf q}) \; \delta n_S(\omega,{\bf q})\delta n_S(-\omega, -{\bf q})
\label{eq:S0}
\end{equation}
where the free  density suceptibility is given by 
\begin{equation}
\chi_S(\omega,{\bf q})=\frac{{\bf v}_S\cdot {\bf q}}{\omega-{\bf v}_S\cdot{\bf q}}
=
\frac{\cos\left(\theta_S-\phi\right)}{s-\cos\left(\theta_S-\phi\right)}
\label{eq:chis}
\end{equation}
with ${\bf v}_S=v_F(\cos\theta_S,\sin\theta_S)$ and ${\bf q}=q(\cos\phi,\sin\phi)$. Thus, $\theta_S-\phi$ is the angle subtended between the momentum ${\bf q}$, and  the Fermi velocity ${\bf v}_S$. In the last term of Eq. (\ref{eq:chis}) we have introduced the dimensionless variable $s=\omega/v_F q$. 

The interacting part of the action is given by, 
\begin{align}
S_I=F_2\int \frac{d^2 qd\omega}{(2\pi)^3}\sum_{S,T}&\cos\left[ 2\left(\theta_S-\theta_T\right)\right]\delta n_T(\omega,{\bf q})\delta n_S(-\omega,-{\bf q})
\end{align}
Note that the free, as well as, the interacting part of the action are quadratic in $\delta n_S$, then, it can be solved exactly. 
However, the calculation is not trivial since, while the free term is diagonal in the patch variable, the interaction term mixes patches.

The main goal of the present calculation is to write the action only in terms of the quadupolar moment, in order to read the corresponding susceptibility. To do this, we note that $\delta n_S$ is a periodic function of $\theta_S$, provided the Fermi surface is closed. Then, we can write the following Fourier expansion, 
\begin{align}
\delta n_S({\bf q})&=m_0({\bf q})+\sum_{\ell=1}^\infty\left\{m_\ell^+({\bf q})\cos\left(\ell\theta_S\right)+
m_\ell^-({\bf q})\sin\left(\ell\theta_S\right)\right\}
\label{eq:deltansmell}
\end{align}
In is not difficult to check  the transformation laws of each component $m_\ell^\pm$. 
Since $\delta n_S$ is a scalar under rotations, it is immediate to see that $m_0$ is also a scalar and is related with the total density of the system.  On the other hand, 
the couple $(m_1^+,m_1^-)$ transform as a vector under rotations and it is related with the total current. Moreover,  $(m_2^+,m_2^-)$, does not transform as a vector but as the components of a symmetric traceless tensor as indicated in Eq. (\ref{eq:Q}), thus $m_2^\pm$ are the components of the quadrupolar moment.  

The interaction term, $S_I$, only contains the quadrupolar components $m_2^\pm$ and it is simply given by
\begin{align}
S_I=\frac{F_2}{2}\int \frac{d^2 qd\omega}{(2\pi)^3} \left\{ \left|m_2^{+}(\omega,{\bf q})\right|^2 + \left|m_2^{-}(\omega,{\bf q})\right|^2\right\}
\end{align}
that, it is evidently rotation invariant, since it is proportional to $Tr(Q^2)$. 

On the other hand, the free part of the action, $S_0$, is not diagonal in Fourier space and consequently mixes different momenta. 
Fortunately, since the interaction only contains $m_2^\pm$, $S_0$ can be diagonalized  and the components $m_\ell^{\pm}$ with $\ell\neq 2$, can be exactly integrated out, providing an action written only in terms of the quadrupolar components $m_2^{\pm}$.
An equivalent procedure is to consider just quadrupolar deformations
\begin{equation}
\delta n_S({\bf q})=m_2^+({\bf q})\cos\left(n\theta_S\right)+
m_2^-({\bf q})\sin\left(n\theta_S\right)
\label{eq:deltans2}
\end{equation}
Replacing Eq. (\ref{eq:deltans2}) into Eq. (\ref{eq:S0}) and integrating over the angles $\theta_S$, we obtain the effective action in terms of the quadrupolar momentum $m_2^\pm$. In a covariant notation we find, 
\begin{align}
S&=\frac{1}{2}\int \frac{d^2q d\omega}{(2\pi)^3}\left\{ \left[\Gamma_0(s)-F_2\right]Tr \hat Q^2-\Gamma_4(s) Tr \left(\hat P\hat Q\right)^2 \right\}
\label{eq:Scovariant} 
\end{align}
where the tensor $\hat Q$ is given by Eq. (\ref{eq:Q}) and we have introduced a second order tensor $\hat P$ given by
\begin{equation}
\hat P=\left(
\begin{array}{cc}
\cos(2\phi)  & \sin(2\phi) \\
 & \\
\sin(2\phi)  & -\cos(2\phi) 
\end{array}
\right)
\end{equation}
The coefficients $\Gamma_0$ and $\Gamma_4$ are given in terms of the free susceptibilies,
\begin{eqnarray}
\Gamma_0(s)&=&\frac{\chi_0^0(s)}{(\chi_0^0(s))^2-(\chi_0^4(s))^2} 
\label{eq:Gamma0} \\
\Gamma_4(s)&=&\frac{\chi_4^0(s)}{(\chi_0^0(s))^2-(\chi_0^4(s))^2} 
\label{eq:Gamma2}
\end{eqnarray}
where,
\begin{equation}
\chi_0^\ell(s)=\int \frac{d\theta}{2\pi}  \frac{\cos\theta}{s-\cos\theta}e^{i\ell \theta}
\end{equation}
Integrating this expression  for $s>1$ and analytically continuing for $s<1$ we find in particular, 
\begin{align}
\chi_0^0(s)&=-1+K_0(s) \\
\chi_0^4(s)&=K_0(s) \left(\frac{1-K_0(s)}{1+K_0(s)}\right)^2
\end{align}
with
\begin{equation}
K_0(s)=\frac{s}{\sqrt{s^2-1}}
\label{eq:K0}
\end{equation}
We can also explicitly re-write the action of Eq. (\ref{eq:Scovariant}) in a non-covariant form, in terms of $m_2^\pm$ finding, 
\begin{align}
S=\int \frac{d^2 qd\omega}{(2\pi)^3} &\left\{ \left(\Gamma_0(s)-\Gamma_4(s)\cos(4\phi)-F_2\right) \left|m_2^{+}\right|^2
+\left(\Gamma_0(s)+\Gamma_4(s)\cos(4\phi)-F_2\right) \left|m_2^{-}\right|^2\right. \nonumber \\
& \left.-\Gamma_4(s)\sin(4\phi) \left[m_2^+m_2^{-*}+m_2^{+*}m_2^-\right]
\right\}
\end{align}
This is the effective action for the quadrupolar moments $m_2^{\pm}(\omega,{\bf q})$. Due to the tensor character of $m_2^\pm$, the action explicitly depends on the angle $\phi$. For arbitrary direction, the action is not diagonal in the components $m_2^\pm$.
We can gain more intuition by fixing a particular direction, for instance ${\bf q}=q\hat i$, {\em i.\ e.\ }, $\phi=0$. In this case, the Lagrangian density has the simpler diagonal form, 
\begin{align}
{\cal L}&=\left(\Gamma_0(s)-\Gamma_4(s)-F_2\right) \left|m_2^{+}(\omega,{\bf q})\right|^2+\left(\Gamma_0(s)+\Gamma_4(s)-F_2\right) \left|m_2^{-}(\omega,{\bf q})\right|^2 
\label{eq:Lagrangian}
\end{align}
from which we can read the longitudinal and transverse inverse quadrupolar susceptibilities
\begin{equation}
\left(\chi_2^\pm\right)^{-1}=\Gamma_0(s)\mp\Gamma_4(s)-F_2
\end{equation}
Inverting this expression and using Eqs. (\ref{eq:Gamma0}) and (\ref{eq:Gamma2}) we finally find, 
\begin{equation}
  \chi_2^{\pm}(\omega,{\bf q})=\frac{\chi^0_0(s) \pm \chi^0_{4}(s)}{
    1-F_{2}\big(\chi^0_0(s) \pm \chi^0_{4}(s)\big)}
\label{eq:chi2}
\end{equation}
This result coincides with that presented in Refs. \onlinecite{OgKiFr2001,Lawler2006} and with Eq. (2) o the main paper.

\section{Effective action near the exceptional point}
The effective Lagrangian of Eq. (\ref{eq:Lagrangian}) is highly non-local in time. In fact, it has a cut for $s<1$. In order to have a simpler expression, let us expand the longitudinal part around the branch point $s=1$. 
We introduce a small variable $z=\sqrt{(s-1)/2}$. In term of $z$, we can simplify $K_0(s)$ given by Eq. (\ref{eq:K0}), 
\begin{equation}
K_0(s)\sim\frac{1}{2}\frac{1}{z}
\end{equation}
Within this approximation
\begin{equation}
\chi_0^0+\chi_0^4\sim\frac{1}{z}-5-4 z+ O(z^2)
\end{equation}
Replacing this expression in the longitudinal part of the Lagrangian we find the kernel
\begin{equation}
{\cal L}^+=\Gamma_0(s)-\Gamma_4(s)-F_2
\sim -F_2+z+5 z^2 + O(z^3)
\end{equation}
Turning back to the original variables we obtain
\begin{equation}
{\cal L}^+(s)=-F_2+\sqrt{\frac{s-1}{2}}+5\left(\frac{ s-1}{2}\right)+O\left((s-1)^{3/2}\right).
\label{eq:L1}
\end{equation}
which coincides with Eq. (4) of the paper. 

We are interested in the zeros of ${\cal L}^+(s)$. Therefore, solving ${\cal L}^+(s)=0$ we obtain
\begin{equation}
s_\pm=\frac{1}{25}\left\{\left(26+10 F_2\right)\pm \sqrt{20 F_2+1}\right\}\; .
\label{eq:spm}
\end{equation} 
Note that the zeros are generally order one, except at $F_2=F_c=-1/20$, where both zeros are degenerate $s_+=s_-=51/50>1$.
For $F_2<F_c$, the zeroes are complex conjugates. 
Since for $F_2>F_c$, $s_\pm>1$, the kernel ${\cal L}^+(s)$ should be an  analytic function of $s$ in a neighborhood of these points. Thus, we can write 
\begin{equation}
{\cal L}^+(s)=\left(s-s_+\right)\left(s-s_-\right)
\label{eq:L2}
\end{equation}
However, it is convenient to re-write this expression in terms of real positive numbers.  Using the explicit expression for $s_\pm$, given by Eq. (\ref{eq:spm}) we easily find, 
\begin{equation}
S_{\rm eff}=\int ds\;  \left\{ \left(s-\epsilon_1\right)\left(s-\epsilon_2\right)+w^2\right\}|m_2^+(s,{\bf q})|^2
\label{eq:S1}
\end{equation}
where we have defined
\begin{align}
\epsilon_1&=\frac{1}{25}\left(27+10F_2\right) \\ 
\epsilon_2&=\frac{1}{25}\left(25+10F_2\right) \\
w&=\frac{1}{25}\sqrt{20 |F_2|}
\end{align}

Therefore, near the singular point $s_+\sim s_-$, the longitudinal component of the quadrupolar momentum satisfy the following 
second order equation 
\begin{equation}
\left\{\left(s-\epsilon_1\right)\left(s-\epsilon_2\right)+w^2\right\} m_2^+(s,{\bf q})=0
\end{equation} 
Near the EP, $s\neq \epsilon_2$, then we can write the equation as 
\begin{equation}
\left\{\left(s-\epsilon_1\right)-i w \left( \frac{iw}{s-\epsilon_2}   \right)\right\} m_2^+(s,{\bf q})=0
\end{equation} 
Introducing the variables 
\begin{align}
\delta Q_1&= m_2^+ \\
\delta Q_2&= \frac{iw}{s-\epsilon_2}  m_2^+ 
\end{align}
we immediately find
\begin{align}
&\left(s-\epsilon_1\right)\delta Q_1 -i w \delta Q_2=0 \\
&\left(s-\epsilon_2\right)\delta Q_2 -i w \delta Q_1=0
\end{align}
These first order system of equations can be obtained from a Hamiltonian formalism. 
By introducing  a two-component vector field $\delta Q^+=(\delta Q_1, \delta Q _2)$ the effective action is written as
\begin{equation}
S_{\rm eff}=\int ds\;  (\delta Q^+)^\dagger\left( s I- H_{\rm eff}\right )\delta Q^+   
\label{eq:Seff}
\end{equation}
where  $I$ is the $2\times 2$ identity matrix and the effective Hamiltonian is 
\begin{equation}
H_{\rm eff}=\left(
\begin{array}{cc}
\epsilon_1 & iw  \\
iw & \epsilon_2
\end{array}
\right)
\end{equation} 
as shown in Eq. (9) of the main paper. 


%

\end{document}